\documentclass[cmex10,reqno]{amsart}

\usepackage{graphicx}      
\usepackage{amssymb}
\providecommand{\eq}{\triangleq}
\DeclareMathOperator*{\argmin}{arg\,min}

\newcommand{\field}[1]{\mathbb{#1}}
\newcommand{\R}{\field{R}}
\newcommand{\N}{\field{N}}

\newcommand{\T}{\top}
\newcommand{\rank}{\mathrm{rank}}
\newcommand{\diag}{\mathrm{diag}}
\newcommand{\hfs}{\hfill\ensuremath{\square}}

\newtheorem{ass}{Assumption}
\newtheorem{rem}{Remark}
\newtheorem{lem}{Lemma}
\newtheorem{thm}{Theorem}

\begin{document}

\title[Sparse Representations for Packetized Predictive Networked Control]{
Sparse Representations for Packetized Predictive Networked Control%
\footnote{The research was supported in part under Australian Research Council's
Discovery Projects funding scheme (project number DP0988601).}
} 


\author[M. Nagahara]{Masaaki Nagahara}
\author[D. E. Quevedo]{Daniel E.~Quevedo}
\address{M. Nagahara is with
	Graduate School of Informatics,
	Kyoto University, JAPAN.
	D. E. Quevedo is with 
	School of Electrical Engineering \& Computer Science,
	The University of Newcastle, AUSTRALIA.
}
\address{The corresponding author is Masaaki Nagahara.
	Mailing address: Graduate School of Informatics, Kyoto University, Yoshida Honmachi, Sakyo-ku, Kyoto 606-8501, Japan.
	}
\email{nagahara@ieee.org}


\maketitle

\begin{abstract}                
We investigate a networked control architecture for  LTI plant
models with a scalar input. Communication from 
controller to actuator is over an unreliable
 network which introduces packet dropouts. To achieve
robustness against dropouts, we adopt a packetized predictive
control paradigm wherein each control packet transmitted contains tentative
future plant 
input values. The novelty of our approach is that we seek that the
control packets transmitted be sparse. For that purpose, we adapt tools from the
area of compressed sensing and propose to design the control packets via on-line
minimization of a suitable
\mbox{$\ell^1\!\slash \ell^2$} cost function. We then show how to choose parameters of the
cost function to ensure that the resultant closed loop system be  practically
stable, provided the maximum number of consecutive packet dropouts is bounded. A
numerical example illustrates that sparsity reduces 
bit-rates, thereby making our proposal suited to control over unreliable and
bit-rate limited networks.  
\end{abstract}

\section{Introduction}
Compressed sensing, which is a hot topic in signal processing,
aims at reconstructing signals by assuming that the original signal is sparse;
see, e.g.,\cite{CanWak08,ZibEla10}. 
The core idea used in this area is to introduce a sparsity index
in the optimization used for decoding. To be more specific,
the sparsity index of a vector $v$ is defined by the amount of
nonzero elements in $v$ and is usually denoted by $\|v\|_0$,
called the ``$\ell^0$ norm.''
The compressed sensing decoding problem is then formulated by
optimization with $\ell^0$-norm regularization.
The associated  optimization problem is however hard to solve, since it is a
combinatorial one. Thus, it is common to introduce a convex
relaxation by replacing the $\ell^0$ norm with the $\ell^1$ norm. 
Under some assumptions, the solution of this relaxed optimization is known
to be exactly the same as that of the $\ell^0$-norm optimization \cite{CanWak08}.
That is, by minimizing the $\ell^1$-norm, one can obtain a sparse solution.

\par The purpose of the present work is to use sparsity-inducing
techniques  in the context of controller design for
  networked control applications. In particular, we will focus on packetized
  predictive control (PPC); see, e.g., 
\cite{Bem98,CasMosPap06,TanSil07,QueNes11}.
As in regular model predictive control (MPC) formulations,
a cost function is used in PPC to design the controller output. 
Each control \emph{packet}  contains a sequence
of tentative plant  inputs for a finite horizon of
future time instants. Packets which are received at
the plant actuator side, are stored in a buffer to be 
 used whenever later packets are dropped by the network. When there are no
 dropouts, PPC reduces to model predictive control. For PPC to give  desirable
 closed loop properties,  the more unreliable the network is, the larger 
the horizon length (and thus the number of tentative plant input values
contained in each packet) should be.
Therefore, to encompass  bit-rate
limitations of digital networks,  it
becomes natural to seek that the control packets provided 
by PPC be sparse.

\par In order to obtain sparse control packets, we propose to design the latter by
minimization of an $\ell^1\!\slash \ell^2$ cost function, which allows one to
trade
control performance for sparsity of the packets. 
The associated optimization can be effectively solved by iteration methods, as
in compressed sensing applications, see also \cite{DauDefMol04,BecTeb09},
and is, thus, suitable for practical control implementations.
 We  show how to choose the parameters of
the cost function to achieve practical stability of the closed loop in the
presence of bounded packet dropouts. We  then illustrate that
sparsity may reduce
bit-rates. This makes  the proposed control method suitable for
situations where the network is not only unreliable, but also 
bit-rate limited. 

\par Before proceeding, we note that only few works on MPC (and none on PPC)
have explicitly studied the use of cost functions with $\ell^1$ norms
\cite{KeeGil88,LazHeeWeiBem06}. In fact, most MPC
formulations use  quadratic cost functions 
(see e.g., \cite{GooSerDon,Mac,RawMay}).
To the best of our knowledge, no specific results on stability
of mixed $\ell^1\!/\ell^2$ MPC (or the, more general, PPC) have been
documented. 
We also note that our approach  for obtaining sparsity  differs somewhat
 from 
that used in  compressed sensing. In fact,
our objective is to derive sparse signals for efficient {\em encoding},
whereas compressed sensing aims at {\em decoding} 
sparse signals. 


The remainder of this work  is organized as follows:
Section~\ref{sec:plant-model-control} revises basic elements of packetized
predictive control. 
In Section \ref{sec:design}, we show how to choose the cost function to obtain
sparse control packets.
In Section \ref{sec:stability}, we study  stability of the resultant networked
control system. A numerical example is included in 
Section \ref{sec:examples}. Section \ref{sec:conclusion} draws conclusions.

\subsubsection*{Notation:}
We write $\N_0$ for $\{0, 1, 2, 3, \ldots\}$, $|\cdot |$ refers to
modulus 
of a
number. The identity
matrix (of appropriate dimensions) is denoted via $I$. 
For a matrix (or a vector) $A$, $A^\T$ denotes the transpose.
For a vector $v=[v_1,\ldots,v_n]^\T\in\R^n$ and a positive definite matrix $P>0$, 
we define
\[
 \|v\|_P := \sqrt{v^\T P v}, \quad \|v\|_1:=\sum_{i=1}^n |v_i|,\quad \|v\|_\infty:= \max_{i=1,\ldots,n} |v_i|
\]
and also denote $\|v\|_2:= \sqrt{v^\T  v}$.
For any  matrix $P$, $\lambda_{\max}(P)$ and $\lambda_{\min}(P)$
 denote the maximum and the minimum eigenvalues of $P$, respectively.
We also define $\sigma_{\max}^2(P):=\lambda_{\max}(P^\T P)$.

\section{Packetized Predictive Networked Control}
\label{sec:plant-model-control}
We consider the following discrete-time linear and time-invariant plant model:
\begin{equation}
  \label{eq:plant}
  x(k+1)=Ax(k)+Bu(k), \quad k\in\N_0, \quad x(0)=x_0,
\end{equation}
where $x(k)\in\R^n$ and $u(k)\in\R$ for $k\in\N_0$.
We assume that the realization $(A,B)$ is reachable.

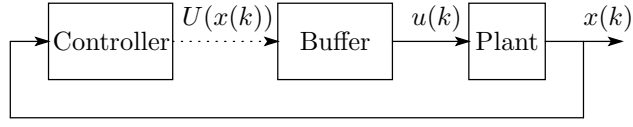
\begin{figure}[tbp]
\begin{center}
\unitlength 0.1in
\begin{picture}( 32.0000,  6.2000)(  2.0000,-10.0000)
%
\special{pn 8}%
\special{pa 400 400}%
\special{pa 1050 400}%
\special{pa 1050 800}%
\special{pa 400 800}%
\special{pa 400 400}%
\special{fp}%
%
\special{pn 8}%
\special{pa 1050 600}%
\special{pa 1600 600}%
\special{dt 0.045}%
\special{sh 1}%
\special{pa 1600 600}%
\special{pa 1534 580}%
\special{pa 1548 600}%
\special{pa 1534 620}%
\special{pa 1600 600}%
\special{fp}%
%
\special{pn 8}%
\special{pa 1600 400}%
\special{pa 2200 400}%
\special{pa 2200 800}%
\special{pa 1600 800}%
\special{pa 1600 400}%
\special{fp}%
%
\special{pn 8}%
\special{pa 2200 600}%
\special{pa 2600 600}%
\special{fp}%
\special{sh 1}%
\special{pa 2600 600}%
\special{pa 2534 580}%
\special{pa 2548 600}%
\special{pa 2534 620}%
\special{pa 2600 600}%
\special{fp}%
%
\special{pn 8}%
\special{pa 2600 400}%
\special{pa 3000 400}%
\special{pa 3000 800}%
\special{pa 2600 800}%
\special{pa 2600 400}%
\special{fp}%
%
\special{pn 8}%
\special{pa 3000 600}%
\special{pa 3400 600}%
\special{fp}%
\special{sh 1}%
\special{pa 3400 600}%
\special{pa 3334 580}%
\special{pa 3348 600}%
\special{pa 3334 620}%
\special{pa 3400 600}%
\special{fp}%
%
\special{pn 8}%
\special{pa 3200 600}%
\special{pa 3200 1000}%
\special{fp}%
\special{pa 3200 1000}%
\special{pa 200 1000}%
\special{fp}%
\special{pa 200 1000}%
\special{pa 200 600}%
\special{fp}%
%
\special{pn 8}%
\special{pa 200 600}%
\special{pa 400 600}%
\special{fp}%
\special{sh 1}%
\special{pa 400 600}%
\special{pa 334 580}%
\special{pa 348 600}%
\special{pa 334 620}%
\special{pa 400 600}%
\special{fp}%
\put(7.3000,-6.0000){\makebox(0,0){Controller}}%
\put(19.0000,-6.0000){\makebox(0,0){Buffer}}%
\put(28.0000,-6.0000){\makebox(0,0){Plant}}%
\put(32.0000,-5.5000){\makebox(0,0)[lb]{$x(k)$}}%
\put(11.0000,-5.5000){\makebox(0,0)[lb]{$U(x(k))$}}%
\put(23.0000,-5.5000){\makebox(0,0)[lb]{$u(k)$}}%
\end{picture}%
\end{center}
\caption{Networked Control System with PPC. The dotted line indicates an erasure
  channel.}
\label{fig:NCS}
\end{figure}

We are interested in a networked control architecture where the controller
communicates with the plant actuator through an unreliable channel, see Fig.~\ref{fig:NCS}.
This channel introduces 
packet-dropouts, which we model via the
dropout sequence $\{d(k)\}_{k \in \N_0}$ in:
\[
d(k)\eq\left\{
\begin{array} {ll}
1, & \textrm{if packet-dropout occurs at instant $k$,}\\
0, & \textrm{if packet-dropout does not occur at instant $k$.}\\
\end{array}\right .
\]

In packetized predictive  control (PPC), as
described, e.g., in \cite{QueNes11,QueOstNes10}, 
 at each time instant $k$, the controller uses the state $x(k)$ of the plant
\eqref{eq:plant} to calculate and
send a control packet of the form
\begin{equation}
  \label{eq:packet}
  U(x(k))\eq\begin{bmatrix}
    u_0(x(k)) &  u_1(x(k))&  \dots & u_{N-1}(x(k))
  \end{bmatrix}^\T\in\R^N
\end{equation}
to the plant input node. 

\par To achieve robustness
against packet dropouts, buffering is used.
To be more specific, suppose that at time instant $k$, we have $d(k)=0$, i.e,
the  
data packet $U(x(k))$ defined in \eqref{eq:packet} is successfully received at
the plant input side. Then, this packet is stored in a buffer, overwriting its
previous contents.
If the next packet $U(x(k+1))$ is dropped,
then the plant input $u(k+1)$ is set to $u_1(x(k))$, the second element of
$U(x(k))$.
The elements of $U(x(k))$ are then successively used until
some packet 
is successfully received, i.e., no dropout occurs.
More formally,  the sequence of  buffer states, say
$\{{b}(k)\}_{k\in\N_0}$, satisfies the recursion
\begin{equation}
  \label{eq:buffer}
  b(k)= d(k) S b(k-1) + (1-d(k))  U(x(k)),
\end{equation}
where $ b(0)=0 \in\R^{N}$ and with
\[
    S\eq
\begin{bmatrix}
    0 & 1& 0 & \ldots & 0\\
    \vdots & \ddots & \ddots &\ddots  & \vdots\\
    0 &  \ldots      &  0 & 1  & 0\\
    0 & \hdotsfor{2} &  0 & 1\\
    0 & \hdotsfor{3} &  0
\end{bmatrix}\in\R^{N\times N}.
\]
The buffer states ultimately give rise to the plant inputs in~\eqref{eq:buffer}
via
$$  u(k)=
  \begin{bmatrix}
    1 & 0 & \dots & 0
  \end{bmatrix}
 {b}(k).$$
\section{Sparse Control Packet Design}
\label{sec:design}
As foreshadowed in the introduction, in the present work we seek that the control
packets $\{ U(x(k))\}_{k\in\N_0}$ be sparse. For that purpose, we propose to
use a dynamic $\ell^1\!\slash \ell^2$ optimization. More 
precisely,   at each time
instant $k$, the controller 
minimizes the following cost function:
\begin{equation}
  \label{eq:cost}
   J(U, x(k)) \eq \|x(N|k)\|_P^2 + \sum_{i=0}^{N-1} \|x(i|k)\|_Q^2
	+ \mu \sum_{i=0}^{N-1}|u_i|,
\end{equation}
where $U=[u_0,u_1,\ldots,u_{N-1}]^\T$. In \eqref{eq:cost},  $\{x(i|k)\}$,
$i\in\{0,1,\dots , N\}$ are predicted plant states, which are calculated
 by
$
 x(i+1|k) = A x(i|k) + B u_i,\quad i=0,1,\ldots,N-1
$
with $x(0|k)=x(k)$, the observed state of the plant \eqref{eq:plant} at time instant $k$.
The parameters $P>0$, $Q>0$, and $\mu>0$ allow the designer to trade control
performance for control effort and sparsity of the control packets. As we will
see in Section~\ref{sec:stability}, the choice of design parameters also
influences closed loop stability of the resultant networked control system.

If we introduce  the following matrices:
\[
 \begin{split}
  \Phi &\eq \begin{bmatrix}B & 0 & \ldots & 0\\
  			AB & B & \ldots & 0\\
			\vdots & \vdots & \ddots & \vdots\\
			A^{N-1}B & A^{N-2}B & \ldots & B
			\end{bmatrix},\quad
  \Upsilon \eq \begin{bmatrix}A\\A^2\\\vdots\\A^N\end{bmatrix},\\
\bar{Q} &\eq \mathrm{blockdiag}\{\underbrace{Q,\ldots,Q}_{N-1}, P\},
 \end{split}
\]
then 
the cost function in \eqref{eq:cost} can be re-written in vector form via 
\begin{equation}
  J(U, x(k)) 
   =\|GU-Hx(k)\|_2^2 + \mu \|U\|_1 + \|x(k)\|_Q^2,
   \label{eq:l1l2}
\end{equation}
where
$G \eq \bar{Q}^{1/2}\Phi,\quad H \eq -\bar{Q}^{1/2}\Upsilon.$
Consequently, the optimal $U(x(k))$ minimizing \eqref{eq:l1l2}
can be numerically obtained by the following iteration \cite{ZibEla10}:
\begin{equation*}
   U(x(k)) \eq \argmin_{U}J(U,x(k)) = \lim_{j\rightarrow\infty} U_{j},
\end{equation*}
where 
\begin{equation}
 U_{j+1} =
 \mathcal{S}_{2\mu/c}\left(\frac{1}{c}G^\T(Hx(k)-GU_j)+U_j\right),\;j\in\N_0,
\label{eq:IST}
\end{equation}
with
\newcommand{\sgn}{\mathrm{sgn}}
\begin{equation}
 \begin{split}
   \mathcal{S}_{2\mu/c}(v) &\eq \begin{bmatrix}\sgn(v_1)\max\{0,|v_1|-2\mu/c\}\\\vdots\\\sgn(v_m)\max\{0,|v_m|-2\mu/c\}\end{bmatrix},
  \end{split}
 \label{eq:IST2}
\end{equation}
see Fig.~\ref{fig:nonlinear}.
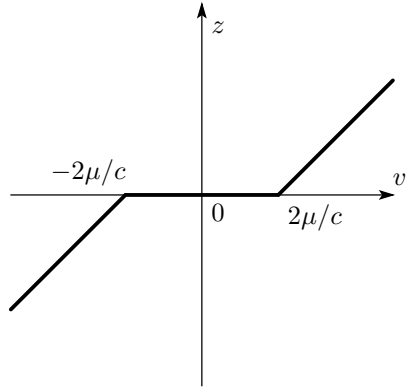
\begin{figure}[tbp]
\begin{center}
\unitlength 0.1in
\begin{picture}( 24.8000, 20.4000)( -0.8000,-20.0000)
%
\special{pn 8}%
\special{pa 400 1000}%
\special{pa 2400 1000}%
\special{fp}%
\special{sh 1}%
\special{pa 2400 1000}%
\special{pa 2334 980}%
\special{pa 2348 1000}%
\special{pa 2334 1020}%
\special{pa 2400 1000}%
\special{fp}%
%
\special{pn 8}%
\special{pa 1400 2000}%
\special{pa 1400 0}%
\special{fp}%
\special{sh 1}%
\special{pa 1400 0}%
\special{pa 1380 68}%
\special{pa 1400 54}%
\special{pa 1420 68}%
\special{pa 1400 0}%
\special{fp}%
%
\special{pn 20}%
\special{pa 400 1600}%
\special{pa 1000 1000}%
\special{fp}%
\special{pa 1000 1000}%
\special{pa 1800 1000}%
\special{fp}%
\special{pa 1800 1000}%
\special{pa 2400 400}%
\special{fp}%
\put(14.5000,-10.5000){\makebox(0,0)[lt]{$0$}}%
\put(18.5000,-10.5000){\makebox(0,0)[lt]{$2\mu/c$}}%
\put(10.0000,-9.5000){\makebox(0,0)[rb]{$-2\mu/c$}}%
\put(24.0000,-9.5000){\makebox(0,0)[lb]{$v$}}%
\put(14.5000,-1.5000){\makebox(0,0)[lb]{$z$}}%
\end{picture}%
\end{center}
\caption{Function $z=\sgn(v)\max\{0,|v|-2\mu/c\}$, $v \in \R$, used
  in \eqref{eq:IST2}.}
\label{fig:nonlinear}
\end{figure}
The constant $c$ in \eqref{eq:IST}  is chosen to satisfy $c>\lambda_{\max}(G^\T
G)$, in which case 
the iteration \eqref{eq:IST} converges to the optimizer of \eqref{eq:l1l2} for any
initial value $U_0\in\R^N$
\cite{DauDefMol04}.
The convergence rate of this iteration is known to be $O(1/j)$. Faster methods
have been proposed, e.g., in \cite{BecTeb09}.

\begin{rem}
  The difference between the cost function in \eqref{eq:cost} and most MPC
  formulations is the penalty on the input vector $U$.  Standard MPC uses an $\ell^2$
  penalty $\|U\|_R^2$, $R>0$ for attenuating the control $U$ (see e.g.,
  \cite{Mac}). In our formulation, we consider an $\ell^1$ penalty $\|U\|_1$,
  which is introduced in order to obtain a sparse representation of the control
  packet $U(x(k))$. \hfill$\square$
\end{rem}

\section{Stability Analysis of $\ell^1\!\slash\ell^2$ PPC}
\label{sec:stability}
We will next  analyze closed loop stability of  $\ell^1\!\slash\ell^2$ PPC,
as presented in Section \ref{sec:design},
with bounded packet dropouts. Our
analysis uses elements 
of the technique introduced in \cite{QueSilGoo07} and which was later refined
in \cite{QueNes11}. 

\par A
distinguishing aspect of the situation at hand is that, for open-loop unstable
plants, even when there are no packet dropouts, asymptotic stability will not be
achieved,  despite the fact that the plant-model in \eqref{eq:plant} is
disturbance-free. This can be easily shown by considering the iteration
in \eqref{eq:IST}, with $U_0=0$ and a plant state 
$x(k)\in\Omega\eq \left\{x\in\R^n: \|G^\T Hx\|_\infty\leq 2\mu\right\}$.
It follows directly from \eqref{eq:IST} that, in this case, 
$\lim_{j\rightarrow\infty}U_j=0$. Since  this limit value 
is independent of the initial value $U_0$,
we have 
$x(k)\in\Omega \Rightarrow U(x(k))=0$.
Thus, if $x(k)\in \Omega$, and there are no dropouts, then
the control $u(k)=0$.
That is, the control system \eqref{eq:plant} behaves as an open-loop system in the set $\Omega$.
Hence, asymptotic stability will in general not be achieved, if $A$
has eigenvalues outside the unit circle. This fundamental property is linked to
sparsity of the control vector.

By the fact mentioned above, we will next study {\em practical} stability (i.e.,
stability of a set) of the
networked control system. For that purpose, we will analyze the  value
function 
\begin{equation}
 V(x) \eq \min_U J(U,x)
 \label{eq:Vx}
\end{equation}
and prepare the following two technical lemmas.:

\begin{lem}[Riccati equation]
\label{lem:ARE}
Let $r>0$.
Suppose 
$P>0$ is the solution to the Riccati equation
\begin{equation}
 P = A^\T PA - A^\T PB (B^\T PB + r)^{-1}B^\T PA + Q,
 \label{eq:ARE}
\end{equation}
and let
$K = -(B^\T PB + r)^{-1}B^\T PA$.
Then
\[
 (A+BK)^\T P(A+BK) - P + Q + rK^\T K = 0.
\]
\end{lem}
{\bf Proof.}
First, we assumed $Q>0$ and hence $(Q^{1/2},A)$ is observable.
We also assumed that $(A,B)$ is reachable. Thus, for any $r>0$ the Riccati
equation has a unique solution $P>0$ 
\cite{ZhoDoyGlo}.
Direct calculation gives the result.\hfs

\begin{lem}[Bounds of $V(x)$]
\label{lem:V_bounds}
For any $x\in\R^n$, we have
\[
  \lambda_{\min}(Q)\|x\|_2^2 \leq V(x) \leq \phi(\|x\|_2)
\]
where
\begin{equation}
 \phi(\|x\|_2)\eq a_1\|x\|_2 + (a_2+\lambda_{\max}(Q))\|x\|_2^2,
 \label{eq:V_bounds}
\end{equation}
\begin{gather*}
  a_1 \eq \mu \sqrt{n}~ \sigma_{\max}\left(G^+ H\right),\quad
  a_2 \eq \sigma_{\max}^2 \left[(GG^+ - I)H\right],\\
  G^+ \eq (G^\T G)^{-1}G^\T.
\end{gather*}
\end{lem}
{\bf Proof.}
First, since $(A,B)$ is assumed to be reachable, we have $B\neq 0$,
and hence $\rank(\Phi) = \rank(\diag(B)) = N$.
This and the assumption that $P>0$ and $Q>0$ imply that
$G=\bar{Q}^{1/2}\Phi$ has full column rank (i.e., $\rank(G) = N$).
Therefore, $G^\T G$ is invertible.
Then consider a vector
\begin{equation}
 \label{eq:1}
 U^\natural(x) \eq G^+ Hx = (G^\T G)^{-1}G^\T Hx.
\end{equation}
Applying this vector to $J(U,x)$ gives
\[
 \begin{split}
  J(U^\natural(x),x) 
   &= \mu \|G^+ Hx\|_1 + \|(GG^+-I)Hx\|_2^2 +  \|x\|_Q^2\\
   &\leq \mu \sqrt{n}\|G^+ Hx\|_2 + \|(GG^+-I)Hx\|_2^2 + \|x\|_Q^2\\
   &\leq \mu \sqrt{n} \sigma_{\max}\left(G^+ H\right)\|x\|_2\\ 
   &\quad+ \left(\sigma_{\max}^2 \left[(GG^+ - I)H\right]+\lambda_{\max}(Q)\right)\|x\|_2^2,
 \end{split}
\] 
where we used the norm inequality $\|v\|_1 \leq \sqrt{n}\|v\|_2$
for any $v\in\R^n$ \cite{Ber}.
This and the definition of $V(x)$ in \eqref{eq:Vx} provide the upper bound on $V(k)$.

To obtain the lower bound, we simply note that by the definition of $J(U,x)$, we have
$\|x\|_Q^2 \leq J(U,x)$
for any $U\in\R^N$, and hence $\lambda_{\min}(Q)\|x\|_2^2\leq V(x)$.
\hfs

\begin{rem}
The vector $U^\natural(x) = G^+ Hx$ used in \eqref{eq:1} is the minimizer of the 
unconstrained cost function
$\|GU-Hx\|_2$ and also approximates $U(x)$, the optimizer of the $\ell^1\!\slash\ell^2$
cost 
\eqref{eq:cost} for plant state $x$. Since the following bound holds \cite{Fuc04}:
\[
  \|G^\T(GU(x) - Hx)\|_\infty = \|G^\T G(U(x) - U^\natural(x))\|_\infty\leq\mu,
\]
the upper bound for $V(x)$ given in Lemma~\ref{lem:V_bounds} will be tight, if
$\mu$ is small. \hfs 
\end{rem}

Having established the above preliminary results, 
we introduce the  iterated mapping $f^{i}$
with implicit (open-loop optimal) input 
\[
U(x)=[u_0(x),\ldots,u_{N-1}(x)]^\T = \argmin_U J(U,x)
\]
by
\begin{equation}
\label{eq:3}
 f^{i}(x) \eq A^ix + \sum_{l=0}^{i-1} A^{i-1-l}Bu_l(x),\quad i=1,2,\ldots, N.
\end{equation}
This mapping describes the plant state evolution during periods of consecutive packet dropouts. 
Note that, since the input $U(x)$ is not a linear function of $x$ (see Section \ref{sec:design}, 
also Fig.~\ref{fig:nonlinear}),
the function $f^{i}(x)$ is nonlinear.
We have the following lemma:
\begin{lem}[Open-loop bound]
\label{lem:lyap}
Assume that $P>0$ satisfies \eqref{eq:ARE} with
\begin{equation}
 r = \frac{\mu^2}{4\epsilon}, \quad \epsilon>0.
 \label{eq:r}
\end{equation}
Then for any $x\in\R^n$, we have
\[
 V(f^{i}(x))-V(x) \leq -\lambda_{\min}(Q) \|x\|_2^2 + \epsilon,\quad i=1,2,\ldots, N.
\]
\end{lem}
{\bf Proof.}
Fix $i\in\{1,\ldots,N-1\}$ and consider the sequence
\[
 \tilde{U} = \left\{u_i(x), u_{i+1}(x), \ldots, u_{N-1}(x), \tilde{u}_N, \ldots, \tilde{u}_{N+i-1}\right\},
\]
where $\tilde{u}_{N+j}$ ($j=0,1,\ldots,i-1$) is given by
\begin{gather*}
  \tilde{u}_{N+j} = K\tilde{x}_{N+j},\quad
  \tilde{x}_{N+j+1} = A\tilde{x}_{N+j} + B \tilde{u}_{N+j},
\end{gather*}
with $K$ as in Lemma~\ref{lem:ARE} and where $\tilde{x}_N = f^N(x)$.
We then have
\[
 \begin{split}
  &J(\tilde{U}, f^{i}(x))\\
   &= \|\tilde{x}_{N+i}\|_P^2
	+ \sum_{l=i}^{N-1} \left\{\|f^l(x)\|_Q^2 + \mu|u_l(x)|\right\}\\
	&\quad + \sum_{l=N}^{N+i-1} \left\{\|\tilde{x}_l\|_Q^2 + \mu|\tilde{u}_l|\right\}\\
   &= V(x) - \sum_{l=0}^{i-1}\left\{\|f^l(x)\|_Q^2 + \mu|u_l(x)|\right\}\\
	&\quad + \|\tilde{x}_{N+i}\|_P^2 - \|f^N(x)\|_P^2
	+ \sum_{l=N}^{N+i-1} \left\{\|\tilde{x}_l\|_Q^2 + \mu|\tilde{u}_l|\right\}\\
   &= V(x) - \sum_{l=0}^{i-1}\left\{\|f^l(x)\|_Q^2 + \mu|u_l(x)|\right\}\\
	&\quad + \sum_{l=N}^{N+i-1} \left\{\|\tilde{x}_{l+1}\|_P^2 - \|\tilde{x}_l\|_P^2 + \|\tilde{x}_l\|_Q^2 + \mu|\tilde{u}_l|\right\}
 \end{split}
\]
By the relation $\tilde{x}_{l+1} = (A+BK)\tilde{x}_l$ and $\tilde{u}_l = K\tilde{x}_l$ for $l=N,N+1,\ldots, N+i-1$,
and by Lemma \ref{lem:ARE}, we can bound the terms in the last sum above by
\[
 \begin{split}
&\|\tilde{x}_{l+1}\|_P^2 - \|\tilde{x}_l\|_P^2 + \|\tilde{x}_l\|_Q^2 + \mu|\tilde{u}_l|\\
  &=\tilde{x}_l^\T\!\left[(A+BK)^\T P(A+BK) - P + Q + \frac{\mu^2 N}{4\epsilon}K^\T K\right]\!\tilde{x}_l\\
	&\quad - \frac{\mu^2N}{4\epsilon}\left(|K\tilde{x}_l|-\frac{2\epsilon}{\mu N}\right)^2 + \frac{\epsilon}{N}
\leq \frac{\epsilon}{N},
 \end{split}
\]
Thus, the cost function $J(\tilde{U}, f^{i}(x))$ can be upper
bounded by
\begin{equation}
 \begin{split}
  J(\tilde{U}, f^{i}(x))
  &\leq V(x) - \sum_{l=0}^{i-1}\left\{\|f^l(x)\|_Q^2 + \mu|u_l(x)| \right\} + \epsilon\\
  &\leq V(x) - \|x\|_Q^2 + \epsilon\\
  &\leq V(x) - \lambda_{\min}(Q) \|x\|_2^2 + \epsilon,
 \end{split}
 \label{eq:J_bound}
\end{equation}
where we have used the relation $f^0(x) = x$.
Since $V(f^i(x))$ is the minimal value of $J(U, f^i(x))$ among all $U$'s in $\R^N$,
we have $V(f^i(x))\leq J(\tilde{U}, f^i(x))$, and hence
\[
 V(f^i(x)) \leq V(x) - \lambda_{\min}(Q) \|x\|_2^2 + \epsilon.
\]
For the case $i=N$, we consider the sequence
$\tilde{U} = \left\{\tilde{u}_N, \tilde{u}_{N+1}, \ldots, \tilde{u}_{2N-1}\right\}$.
If we define $\sum_{l=N}^{N-1} = 0$,
then \eqref{eq:J_bound} follows as in the case $i\leq N-1$.\hfs

The above result can be used to derive the following contraction property of the
optimal costs during periods of successive packet dropouts:
\begin{lem}[Contracting property]
\label{lem:contracting}
Let $\epsilon>0$. Assume that $P>0$ satisfies \eqref{eq:ARE} with $r$ as in \eqref{eq:r}.
Then there exists a real number $\rho\in (0,1)$ such that
for all $x\in\R^n$, we have
\[
 V\left(f^i(x)\right) \leq \rho V(x) + \epsilon + \frac{N\lambda_{\min}(Q)}{4},\quad i=1,2,\ldots, N.
\] 
\end{lem}
{\bf Proof.}
In this proof, we borrow a technique used in the proof of \cite[Theorem 4.2.5]{Laz}.
By Lemma \ref{lem:V_bounds}, for $x\neq 0$ we have
$0 < V(x) \leq a_1 \|x\|_2 + (a_2 + \lambda_{\max}(Q)) \|x\|_2^2$.
\par Now suppose that $0<\|x\|_2\leq 1$.
Then $\|x\|_2^2 \leq \|x\|_2$ and hence
\[
 V(x) \leq (a_1 + a_2 + \lambda_{\max}(Q)) \|x\|_2.
\]
From Lemma \ref{lem:lyap}, it follows that
\[
 \begin{split}
  &V\left(f^i(x)\right) \leq V(x) - \lambda_{\min}(Q)\|x\|_2^2 + \epsilon\\
  &\leq \left(1-\frac{\lambda_{\min}(Q)\|x\|_2}{V(x)}\right)\!V(x)
	\!-\!\lambda_{\min}(Q)\!\left(\|x\|^2_2-\|x\|_2\right) + \epsilon\\
  &\leq \left(1-\frac{\lambda_{\min}(Q)}{a_1+a_2+\lambda_{\max}(Q)}\right)V(x)
	+ \frac{\lambda_{\min}(Q)}{4} + \epsilon\\
  &=\rho V(x) + \frac{\lambda_{\min}(Q)}{4} + \epsilon,
 \end{split}
\]
with
\begin{equation}
 \rho \eq 1-\frac{\lambda_{\min}(Q)}{a_1+a_2+\lambda_{\max}(Q)}.
 \label{eq:rho}
\end{equation}
Since $0<\lambda_{\min}(Q)\leq\lambda_{\max}(Q)$, $a_1>0$, and $a_2>0$,
it follows that $\rho \in (0,1)$.

\par Next,  consider the case where $\|x\|_2>1$ so that
 $\|x\|_2 < \|x\|_2^2$ and
\[
 V(x) < (a_1 + a_2 + \lambda_{\max}(Q))\|x\|_2^2.
\]
This and  Lemma \ref{lem:lyap} give
\[
 \begin{split}
  V\left(f^i(x)\right) &\leq V(x) - \lambda_{\min}(Q)\|x\|_2^2 + \epsilon\\
   &=\left(1-\frac{\lambda_{\min}(Q)\|x\|_2^2}{V(x)}\right)V(x) + \epsilon\\
   &<\left(1-\frac{\lambda_{\min}(Q)}{a_1+a_2+\lambda_{\max}(Q)}\right)V(x) + \epsilon\\
   &=\rho V(x) + \epsilon
   \leq \rho V(x) + \frac{\lambda_{\min}(Q)}{4} + \epsilon.
 \end{split}  
\]
Finally, if $x=0$, then the above inequality also holds since $V(0)=0$.\hfs

We will next use Lemma~\ref{lem:contracting} to establish sufficient conditions
for 
practical
stability of PPC
in the presence of packet-dropouts.  To state our results, in the sequel we
denote the
time instants where there are no
packet-dropouts, i.e., where
$d(k)=0$, as
\[
  \mathcal{K}=\{k_i\}_{i\in\N_0}\subseteq \N_0, \quad
  k_{i+1}>k_i,\; \forall i \in \N_0
\]
whereas the number of consecutive packet-dropouts is denoted via:
\begin{equation}
  m_i\eq k_{i+1}-k_i-1,\quad i\in\N_0.
  \label{eq:mi}
\end{equation}
Note that $m_i \geq 0$,
with equality if and only if no dropouts occur between instants $k_i$ and
$k_{i+1}$.

\par  When packets are
lost, the control system operates in open-loop. Thus, 
to ensure desirable properties of the networked control system, 
one would like the number of consecutive packet-dropouts to be bounded. 
In particular, to establish practical stability,  
we  make the following assumption:%
\footnote{If only stochastic properties are sought, 
then more relaxed assumptions can be used, see related work in \cite{QueOstNes10}.} 

\begin{ass}[Packet-dropouts]
\label{ass:dropouts}
The number of
consecutive packet-dropouts is  uniformly bounded by the prediction horizon
minus one,
i.e., we have $m_i \leq N-1$,  $\forall i \in \N_0$.\hfill$\square$
\end{ass}

Theorem~\ref{thm:stability} stated below shows how to design the parameters of
the cost function to ensure practical stability of the networked control system
in the presence of bounded packet dropouts.

\begin{thm}[Practical stability of $\ell^1\!\slash\ell^2$ PPC]
\label{thm:stability}
Suppose that Assumption \ref{ass:dropouts} holds.
Let $\epsilon>0$ and choose $P>0$ to satisfy \eqref{eq:ARE} with $r$ as in
\eqref{eq:r}. 
Then for all $k\in\N_0$
we have
\begin{equation}
  \|x(k)\|_2 \leq (\sqrt{\rho})^{i+1} \sqrt{\frac{\phi(\|x(k_0)\|_2)}{\lambda_{\min}(Q)}} + \Delta,
  \label{eq:thm}
\end{equation}
where $i\in\N_0$ is such that $k\in\{k_i+1,\ldots,k_{i+1}\}$,
\begin{equation}
  \label{eq:2}
  \Delta \eq \sqrt{\frac{\rho}{1-\rho}\left(\frac{\epsilon}{\lambda_{\min}(Q)}+\frac{N}{4}\right)}
\end{equation}
 $\rho$ is given in
\eqref{eq:rho},  and $\phi(\cdot)$ is as in \eqref{eq:V_bounds}.
\end{thm}

{\bf Proof.}
Fix $i\in\N_0$ and note that at time instant $k_i$, the control packet is successfully transmitted to the buffer.
Then until the next packet is received at time $k_{i+1}$, $m_i$ consecutive packet-dropouts occur.
By the PPC strategy,  the control input becomes 
$u(k_i+l) = u_l(x(k_i))$, $l=1,2,\ldots,m_i$,
and the states $x(k_i+1),\ldots,x(k_i+m_i)$
are determined by these open-loop controls.
Since, by assumption, we have $m_i\leq N-1$,
Lemma \ref{lem:contracting} gives
\begin{equation}
  V(x(k)) 
   \leq \rho V(x(k_i)) + \epsilon + \frac{N\lambda_{\min}(Q)}{4}
   \label{eq:thm1}
\end{equation} 
for $k\in\{k_i+1, k_i+2,\ldots, k_i+m_i\}$, and also for $k_{i+1}=k_i+m_i+1$, we have
\begin{equation}
 V(x(k_{i+1})) \leq \rho V(x(k_i)) + \epsilon + \frac{N\lambda_{\min}(Q)}{4}.
 \label{eq:thm2}
\end{equation}
Now by induction from \eqref{eq:thm2}, it is easy to see that
\[
 \begin{split}
  &V(x(k_i))\\
   &\leq \rho^i V(x(k_0))
	+ (1+\rho+\cdots+\rho^{i-1})\left(\epsilon + \frac{N\lambda_{\min}{Q}}{4}\right)\\
   &\leq \rho^i \phi(\|x(k_0)\|_2) + \frac{1}{1-\rho} \left(\epsilon + \frac{N\lambda_{\min}{Q}}{4}\right).
 \end{split}
\]
This inequality and \eqref{eq:thm1} give the bound
\[
 \begin{split}
  V(x(k)) 
   &\leq \rho^{i+1} \phi(\|x(k_0)\|_2) + \frac{\rho}{1-\rho} \left(\epsilon + \frac{N\lambda_{\min}{Q}}{4}\right)\\
 \end{split}
\]
for $k\in\{k_i+1, k_i+2,\ldots, k_{i+1}-1\}$, and this inequality also holds for $k=k_{i+1}$.
Finally, by using the lower bound of $V(x)$ provided in Lemma \ref{lem:V_bounds}, we have
\[
 \begin{split}
  \|x(k)\|_2
   &\leq \sqrt{\frac{V(x(k))}{\lambda_{\min}(Q)}}
   \leq (\sqrt{\rho})^{i+1} \sqrt{\frac{\phi(\|x(k_0)\|_2)}{\lambda_{\min}(Q)}}
   + \Delta,
 \end{split}
\]
since $\sqrt{a+b} \leq \sqrt{a} +\sqrt{b}$, for all $a,b\geq 0$.\hfs

Theorem \ref{thm:stability} establishes practical stability of the networked
control 
system. It shows that, provided the conditions are met, the plant state will be
ultimately bounded in a  ball of radius
$\Delta$. 
It is worth noting that, as in other stability results which use Lyapunov
techniques, this bound will, in general, not be tight.

\section{Design Examples}
\label{sec:examples}
To illustrate properties of the $\ell^1\!\slash \ell^2$ PPC strategy proposed in
this work, we consider a plant
model of the form \eqref{eq:plant} with%
\footnote{The elements of these matrices are generated by random sampling from
the normal distribution with mean 0 and variance 1.
Note that the matrix $A$ has 3 unstable eigenvalues and 1 stable eigenvalue.}
\[
 \begin{split}
 A &= \left[\begin{array}{rrrr}
     1.2597574&  - 0.265722&   - 0.6776537&    1.1712147\\
  - 0.0066489&  - 0.846387&   - 0.4174316&    1.1930255\\
  - 0.4610984&  - 0.1307435&  - 0.1483141&    0.2842062\\
  - 0.4855527&    0.2480541&    1.8002141&    0.7398921\\
 \end{array}\right],\\
 B &= \left[\begin{array}{r}
     1.3372142\\
  - 2.9903216\\
    0.9703207\\
  - 0.4056704\\
 \end{array}\right].
 \end{split}
\]

We set the  horizon length in the cost function \eqref{eq:cost} to $N=5$ and
choose weights $\mu=100$,
$Q=I$ 
and $P$ as the solution to the Riccati equation \eqref{eq:ARE}
with $r=\mu$. In this case, the real number $\epsilon$ in Theorem \ref{thm:stability}
equals 25, computed by \eqref{eq:r}.

\par For comparison, in addition to the $\ell^1\!\slash\ell^2$ PPC, we also
synthesize PPC with a conventional $\ell^2$ cost function, namely
\begin{equation}
\label{eq:quadratic}
\begin{split}
  J_2(U, x(k)) &\eq \|x(N|k)\|_P^2 + \sum_{i=0}^{N-1} \|x(i|k)\|_Q^2+
	\mu \sum_{i=0}^{N-1}|u_i|^2.
 \end{split}
\end{equation}
We first simulate a
packetized predictive networked control system setup as in
Fig.~\ref{fig:NCS}. The number $m_i$ of consecutive packet dropouts, see
\eqref{eq:mi}, is chosen from the uniform distribution on $\{1,2,3,4\}$. 
The initial plant state is set to $x(0)=[1,1,1,1]^\T$.
Table \ref{tbl:packet} shows the first 5 successfully transmitted control packets $U(k_i)$,
$i=0,1,2,3,4$ designed by the proposed method ($\ell^1\!\slash \ell^2$)
and the quadratic one given in~\eqref{eq:quadratic}. It can be observed that,
for the present situation, the control packets provided by the
$\ell^1\!\slash \ell^2$ design are more sparse than those obtained through the
$\ell^2$ formulation.

\begin{table}[tbp]
\caption{Control packets $U(k_i)$}
\label{tbl:packet}
\begin{center}
\begin{tabular}{|c||r|r|r|r|r|}\hline
  & $i=0$ & $i=1$ & $i=2$ & $i=3$ & $i=4$\\\hline
& - 2.632&    0&      - 1.809&  - 0.085&  - 0.909\\
&   0.085&  - 1.825&    0&      - 0.890&    0\\
$\ell^1\!\slash \ell^2$&  - 2.211&  - 0.022&  - 0.826&    0.21&     0.157\\
&    0&      - 0.753&    0&       0&       0.322\\
&    0&        0&        0&        0&        0\\\hline
& - 2.632&    0.007&  - 1.733&  - 0.137&  - 0.651\\
&  - 0.106&  - 1.74&   - 0.154&  - 0.759&    0.292\\
$\ell^2$&  - 1.869&  - 0.162&  - 0.778&    0.169&  - 0.465\\
&    0.102&  - 0.762&    0.207&  - 0.213&    0.150\\
&  - 0.679&    0.213&  - 0.201&    0.229&  - 0.224\\\hline
\end{tabular}
\end{center}
\end{table}

\begin{figure}[tbp]
\begin{center}
\includegraphics[width=.8\linewidth]{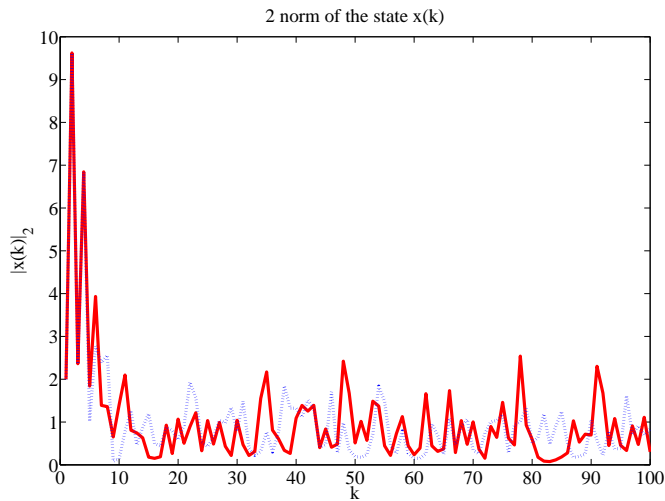}

\end{center}
\caption{2 norm of the state $x(k)$: proposed $\ell^1/\ell^2$ (solid) and conventional $\ell^2$ (dots)}
\label{fig:states}
\end{figure}

\begin{figure}[tbp]
\begin{center}
\includegraphics[width=.8\linewidth]{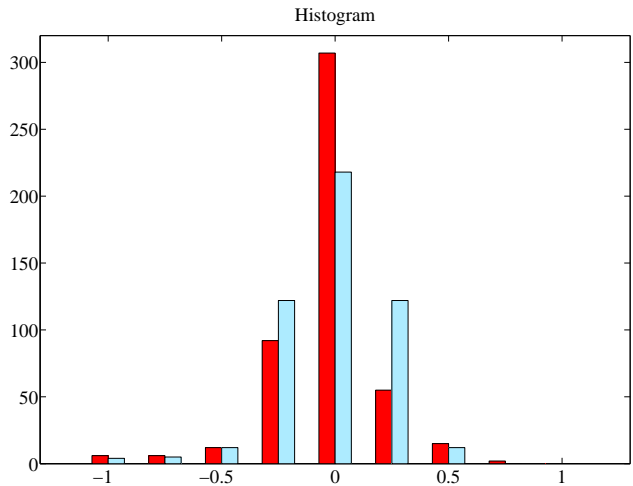}
\end{center}
\caption{Histogram of the quantized values in all transmitted packets: proposed $\ell^1\!\slash \ell^2$ (left) and conventional $\ell^2$ (right)}
\label{fig:histogram}
\end{figure}

To study bit-rate aspects, we next quantize the packets $\{U(k)\}$, by using 
 8-bit uniform quantization with step size 0.25 in each component.
Fig.~\ref{fig:states} illustrates the 2 norm of the state $x(k)$ obtained when using
the $\ell^1\!\slash \ell^2$ PPC and also with the $\ell^2$ PPC.
Whilst both controllers give a loop which seems practically stable and exhibits
 comparable performance,  there is a difference in the quantized packets
 generated by each method. This can be appreciated in Fig.~\ref{fig:histogram},
 which shows a histogram of the quantized control values
$\{u_l(k)\}$, $l\in\{0,1,\ldots,N-1\}$, $k\in\{1,2,\dots,100\}$
in all transmitted packets $U(k)=[u_0(k),\ldots,u_{N-1}(k)]^\top$.
 As indicated in
Fig.~\ref{fig:histogram}, control packets designed by the proposed method include much more zeros
than those obtained  by the conventional one, see~\eqref{eq:quadratic}.
In fact, the packets obtained by $\ell^1\!\slash \ell^2$ optimization include 307 zero
values, whereas the conventional ones contain only 218 zeros.
Since the total amount of transmitted packets is 100, in the $\ell^1\!\slash
\ell^2$ formulation about $3/5$ of the
transmitted signals 
are zero. This observation suggests that the associated bit-rates of the signal
 will be small.


To further investigate this issue, 
we compute the (discrete) entropy  of the control values $\{u_l(k)\}$, defined as
$\mathcal{H}(u) \eq -\sum_{u}\mathcal{P}(u)\log_2 \mathcal{P}(u)$
[bits],
where $\mathcal{P}(u)$ is the probability mass function of $u$; see, e.g., \cite{CovTho}. 
The function $\mathcal{P}(u)$ can be approximately estimated 
by the histogram as in Fig.~\ref{fig:histogram}.
Note that here we do not take account of the entropy of control vector $U(k)$,
that is, we assume scalar quantization.
In the situation studied with $x(0)=[1,1,1,1]^\top$,
the entropy
by the  $\ell^1\!\slash \ell^2$ optimization proposed is
8.6177,
while that by the conventional $\ell^2$ approach  is
9.5345.
We next execute 10000 simulations with initial plant states $x(0)$
whose elements are randomly sampled from the Gaussian distribution with mean 0
and variance 1. The average entropies obtained are
12.2560 for $\ell^1\!\slash \ell^2$ optimization and 15.5701 for the $\ell^2$ formulation.
Since the entropy of the signal transmitted serves as a measure of the code
length,   in the cases studied, the $\ell^1\!\slash \ell^2$  PPC method  will require
 lower bit-rates than PPC with an $\ell^2$  cost function.

\par
\begin{figure}[tbp]
\begin{center}
\includegraphics[width=.8\linewidth]{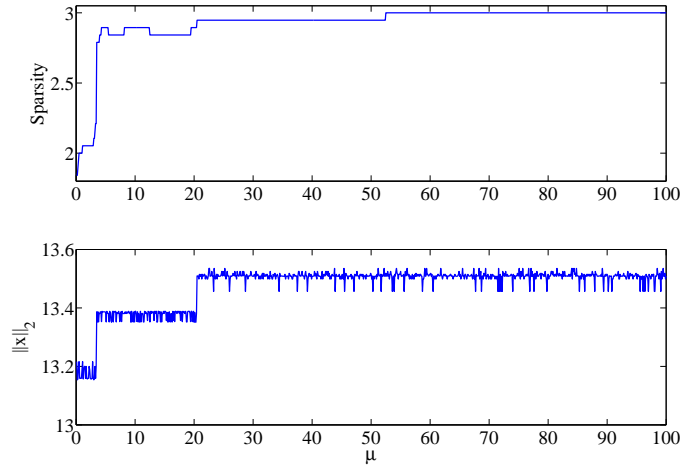}
\end{center}
\caption{Sparsity $N-\|U\|_0$ (top figure) and  performance $\|x\|_2$ (below), as a function
  of $\mu$.}
\label{fig:mu}
\end{figure}

 Fig.~\ref{fig:mu} illustrates the tradeoff between control performance and
sparsity, which, as noted above, is related to bit-rates.
The figure illustrates the average sparsity $5-\|U\|_0$, where
\[
 \|U\|_0 \eq \frac{1}{100}\sum_{k=1}^{100} \|U(k)\|_0,
\]
and the achieved performance $\|x\|_2$ (the 2 norm of the state sequence $x$), for 
 $\mu\in[0,100]$. 
The other parameters are the same as used above.
As can be appreciated, as $\mu$ becomes larger, the sparsity increases,
but the performance becomes worse. 

\section{Conclusion}
\label{sec:conclusion}
We have studied a packetized
predictive control formulation  with  an $\ell^1\!\slash \ell^2$  cost
function. The associated optimization can be solved   effectively
and is, thus, suitable for implementation in a real-time controller.
We have given sufficient conditions for practical stability when the
controller is used over a network with bounded packet dropouts.
Numerical results indicate that the proposed controller provides
 sparse control packets, thereby giving bit-rate reductions
when compared to the use of, more common, quadratic cost  functions.
Future work may include the further study of performance aspects and the effect of
plant disturbances.  

\section*{Acknowledgement}
The authors would like to thank Jan {\O}stergaard for his helpful comments.

\end{document}